\documentclass[useAMS,usenatbib]{mn2e}
\usepackage{graphicx}
\usepackage{latexsym}
\begin{document}

\title[Interaction of winds in binary system PSR~1259-63/SS2883]{Modeling interaction of relativistic and nonrelativistic winds in binary system 
PSR~1259-63/SS2883. I.~Hydrodynamical limit}

\author[Bogovalov et al.]
       {Bogovalov${}^1$ S.V., Khangulyan${}^2$ D., Koldoba${}^3$ A.V., Ustyugova${}^3$ G.V., Aharonian${}^{4,2}$ F.A.\\
       ${}^1$Moscow Engineering Physics Institute (state university), Moscow, Russia,\\
       ${}^2$Max-Planck Institute fur Kernphysik, Heidelberg, Germany\\
       ${}^3$Keldysh Institute of Applied Mathematics RAS, Moscow, Russia,\\
       ${}^4$Dublin Institute for Advanced Studies, Dublin, Ireland.}
\date{ }
\pubyear{2007} 

\maketitle

\begin{abstract}
In this paper, we present a detailed hydrodynamical study of the properties of the flow produced by the collision of a pulsar wind with the surrounding in a binary system. This work is the first attempt to simulate interaction of the ultrarelativistic flow (pulsar wind) with the nonrelativistic stellar wind. Obtained results show that the wind collision could result in the formation of an "unclosed" (at spatial scales comparable to the binary system size) pulsar wind termination shock even when the stellar wind  ram pressure exceeds significantly the pulsar wind kinetical pressure. Moreover, the post-shock flow propagates in a rather narrow region, with very high bulk Lorentz factor ($\gamma\sim100$). This flow acceleration is related to adiabatical losses, which are purely hydrodynamical effects. Interestingly, in this particular case, no magnetic field is required for formation of the ultrarelativistic bulk outflow. The obtained results provide a new interpretation for the orbital variability of radio, X-ray and gamma-ray signals detected from binary pulsar system PSR~1259-63/SS2883.
\end{abstract}
\section{Introduction}

Pulsars lose their rotation energy through  relativistic winds, the collision of which with the Interstellar Medium results 
in the formation of the  Pulsar Wind Nebulae (regions 
of nonthermal synchrotron radiation of ultrarelativistic 
electrons accelerated at the termination of  pulsar winds  \citep{rees74,kc}).
The Crab Nebula is the most famous example of such an object. The recent 
X-ray and TeV gamma-ray observations \citep{GenSlane} show that 
this is a common phenomenon.
 
A very interesting situation arises when a pulsar is located in 
a binary system. In this case the pulsar wind interacts with the 
wind from the companion star.  This case, in particular, 
is realized in the binary system PSR1259-63/SS2883 which consists 
of a  $\sim 48 \rm ms$ pulsar in an elliptic 
orbit around a massive B2e optical companion \citep{johnston1}. 
The density and  velocity of the stellar wind depend on the 
separation distance between two stars. Thus, the processes 
related to the interaction of two winds, 
in particular particle acceleration and radiation proceed 
under essentially different physical conditions depending on the orbital phase.
This makes this object a unique laboratory for the study of nonthermal processes 
in ``on-line'' regime, due to the short 
acceleration and cooling time-scales characterizing this system, especially 
close to the periastron \citep{khangoulian_psr}. Observations show   
that this system is indeed a strong source of nonthermal time-dependent emission 
extending from radio to TeV gamma-rays (see e.g. \cite{neronov}).

Two variable TeV galactic gamma-ray sources, 
LS 5039 and LSI~61~303 (see e.g. \cite{paredes_rev}), 
are discussed as possible, although less evident, candidates 
representing ``binary pulsar'' source population \citep{dubus,mirabel,neronov2}. 
LS 5039 consists of O6.5V star and an unidentified compact object
in a 3.9 day orbit.  This object has been detected as the source of 
gamma rays by EGRET (source 3EG J1824-1514) 
\citep{ls5039EGRET} and by HESS \citep{ls5039}. 
LSI~61~303 is a binary system with a B0Ve star in a 26.5 day orbit. 
This source presumably associates with a low energy gamma-ray source 
2CG 135+01/3EG J0241+6103 \citep{lsiaspulsar,tavani_lsi}.
Recently, it has been 
detected in TeV gamma-rays  \citep{lsi61303}.   
The nature of the compact objects (black hole or a neutron star) 
in both sources is not yet firmly established.

The collision of supersonic winds from two stars located in a binary system 
results in the formation of two terminating shock fronts and a tangential
discontinuity separating relativistic and nonrelativistic parts of the 
shocked flow. It is believed that the shocked flow should propagate 
into a limited solid angle with rather high velocity. However 
the properties  of the flow downstream the shock  is unknown.    

The collision of the pulsar wind with the supersonic flow of the nonrelativistic 
plasma has been recently numerically modeled
by several groups \citep{bucciantini,swaluw,vigelius,romero}
using different nonrelativistic versions of hydrodynamical codes. 
However,  the dynamics of the relativistic plasma is rather specific, and 
the use of nonrelativistic codes cannot be \textit{a priori} justified. 
The  modeling of  collision of a relativistic pulsar wind 
with a nonrelativistic one demands an adequate treatment of distinct features 
of relativistic outflows.  Moreover, in many previous studies the stellar wind has been approximated as 
plane parallel, while both winds initially expand radially. However, this purely geometric 
difference in the formulation of the problem, 
leads, in fact, to significantly different results and conclusions.

In this paper we present the results of our studies conducted in the hydrodynamical limit, 
i.e. the role of the magnetic field in dynamics has been ignored.
The impact of the magnetic field will be published elswhere.      

\section{Properties of the pulsar and stellar winds of PSR 1259-63/SS2883}

Below  we describe the properties of 
PSR 1259-63/SS2883 in which the interaction of the pulsar and stellar winds 
leads to the nonthermal emission observed in radio, X-ray and TeV gamma-ray energy bands. 

\subsection{Pulsar wind}

The rotational losses of pulsars are released in the form of 
electromagnetic and kinetic energy fluxes of the wind. 
Interpretation of observations usually results in a 
conclusion that the energy flux in the wind at large distances 
from the pulsar is concentrated in the particle kinetic energy.  The ratio
of the electromagnetic energy flux over the kinetic energy flux is  described using 
$\sigma$-parameter. Typically  $\sigma$ is much less than 1.  Therefore, it is natural in the first approximation to consider the pulsar wind as purely hydrodynamical. Below we assume 
that the wind is ejected isotropically with the total energy flux equal to the total rotational losses of the pulsar $\dot E_{rot}$. Thus the momentum flux density of the wind varies with 
the distance $r$ to the pulsar  as 
\begin{equation}
L={\dot E_{rot}\over 4\pi c r^2} \ .
\end{equation}
For a given Lorentz factor of the wind $\Gamma_0$, 
the proper density of the plasma in the pre-shock region is equal to  
\begin{equation}
n={\dot E_{rot}\over 4\pi mc^3 \Gamma_0^2 r^2} \ .
\end{equation}

\subsection{Parameters of the stellar wind}

In the system PSR 1259-63/SS2883 the 47.8 ms radio pulsar rotates around the Be 
star in an elliptic orbit with a period 3.4 yr.   
The eccentricity of the orbit is 0.87. The distance between stars in apastron is $\sim 10^{14} \rm cm$ (Johnston et al, 1994). 
The Be star has a mass $\sim 10 \rm M_\odot$ and  radius $R_*=6 R_\odot$.  Due to the 
very fast rotation of the star, the mass outflow  is strongly anisotropic.  The mass flux is concentrated along the equatorial plane forming a disk-like flow. 
In addition, there is also an isotropic component of the wind  
(so-called \textit{polar wind})  with smaller mass flux, but higher velocity. 

The disks in Be stars are dense and expand rather slowly.  
Typical velocities of the plasma in the disk are $150-300 \ \rm km/s$
\citep{waters}. 
The velocity of the plasma in the polar wind is much higher achieving 
$1500-2000 \ \rm km/s$ \citep{snow}.  
The ratio of the mass flux density in the disk  over the polar wind is estimated \citep{waters} as 
 \begin{equation}
{n_d v_d\over n_pv_p}= 30-100 \ .
\end{equation}

The mass flux of the disk-like wind from Be SS2883 has been estimated by several authors.  The estimates were made basically using eclipses by the disk-like flow of the pulsed radio emission from the pulsar. 
Following \cite{johnston}  the mass loss rate 
$\dot M=5\times 10^{-8} \rm M_\odot \ \rm yr^{-1}$ and the 
opening angle of the disk $\theta \sim 15^\circ$. 
Then the ratio of the momentum flux-densities from the Be star and the pulsar, $\eta$, in the case of pulsar--disk interaction, is    
\begin{equation} 
 \eta={E_{rot}\sin(\theta/2)\over c\dot M v} \ . \label{eq_eta}
\end{equation}   

For the given parameters characterizing PSR 1259-63/SS2883, 
$E_{\rm rot}= 8\times 10^{35} \rm erg s^{-1}$, $\theta=15^\circ$, 
and the velocity of the equatorial outflow $v=200\ \rm km/s$, one  obtains
$\eta \simeq 5.5\times 10^{-2}$ with an  uncertainty by a factor of 2-3.  
In the polar wind the velocity is much larger, $v\sim 2000 \ \rm km/s$, but, 
at the same time  the mass-flux is  
reduced by a factor of 30-100. 
Therefore, in the polar wind of the Be star, $\eta$ appears in the range 0.2 - 0.6.  
Thus, given the uncertainties in the mass-flux from the Be  star 
and the variation of the momentum flux with distance due to the acceleration of 
the stellar wind by radiation pressure,
the parameter  $\eta$ along the pulsar orbit may vary  
between $10^{-2}$ and  1.

\section{Computational Methods}

\begin{figure}
\centerline{\includegraphics[width=0.5\textwidth, bb = 0 0 529 302]{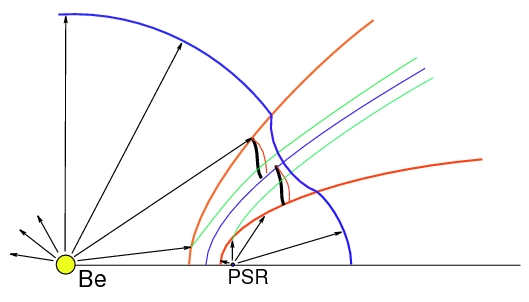}}
\caption{The schematic representation of the flow structure at the collision of the winds. The red lines show the termination shock waves. The thin blue line shows the contact discontinuity. The green lines show typical flow lines in the post shock region. The thick blue line -  the starting surface for the extention of the solution in the hyperbolic region. The thick black lines - sound surfaces. The red lines attached to them - separatrix characteristics.
}
\label{scheme}
\end{figure}

The scheme of the plasma flow formed at the collision of the pulsar wind with the wind from the Be  star is shown in Fig.~\ref{scheme}.  Both winds are terminated with formation of two shock waves.  The post shock flow of the nonrelativistic plasma is separated from the shocked relativistic plasma flow by a  contact discontinuity. Within the hydrodynamical classification,
the contact discontinuity is treated as tangential discontinuity  
\citep{gasodynamica}.  

A remarkable feature of the post-shock flow arising from the collision of the two supersonic winds is that the flow is subsonic only in the region close to the line connecting  the stars (i.e. the symmetry line).  Propagating downstream  the flow crosses \textit{sound line} (shown in Fig.~\ref{scheme} by the thick black lines) and becomes supersonic. We note that the locations  of the \textit{sound lines}  in the nonrelativistic and relativistic flows are different.   While the flow in the subsonic region is described by elliptic-type equations, in the supersonic region the flow is   hyperbolic. Thus the plasma flow in the post shock region appears to be 
of mixed-type.  The \textit{critical lines} or \textit{separatrix characteristics}  play a crucial role in the mixed-type hydrodynamical \citep{gasodynamica,transonic} and magnetohydrodynamical \citep{bogovalov1,tsinganos} flows.
The separatrix characteristics are located in the hyperbolic region, and in Fig.~\ref{scheme} they are shown by thin red lines. The separatrix characteristics and sound lines may coincide only in a particular case when they are orthogonal to the flow lines.  The separatrix characteristics separate the flow regions with different causal connection.  In the ``subsonic'' region limited by the separatrix characteristics every couple of points are causally connected through hydrodynamical 
or MHD signals. Downstream the separatrix characteristics  any two points
are causally disconnected unless they are located in the cone formed by  two different characteristics. Moreover, in a case of steady state mixed type flow,  the solution setting on the separatrix characteristics specifies a unique solution in the hyperbolic region.

The most general numerical method  of mixed-type problem solution  is the relaxation method.  In this method, the time dependent problem of the plasma flow is solved starting from some initial state.  The method
was used for the mixed-type equation solution by many groups in the context of the problem of astrophysical jet formation \citep{BT,romanova,pudrits,krasnopolsky}.  However, the applicability of this method is rather 
limited. Usually transient equations are solved in a rather small computational domain.
Typically, the outer boundaries of the domain are located in the region of the hyperbolic flow.  Theoretically they can be located at arbitrarily large distances  downstream of the critical surfaces. However, in practice  they are located not very far away due to the limitations imposed by memory and power of computers.  At the same time,  one is often interested  in the  plasma flow rather far downstream  from the critical lines. Therefore, it was proposed by \cite{BT}  to  extend  the solution obtained by the relaxation method  to arbitrarily large distances downstream of the separatrix characteristics.  This extention is based on the fact that in the hyperbolic region a Cauchy problem  can be formulated with the initial  data specified at some 
line located slightly downstream of the separatrix characteristic.  Thus, the solution of the problem of the mixed-type flow  can be  reduced to a two-step treatment. 
At the first step (nearest zone solution) the solution 
can be found in the limited computational domain using the relaxation method.  At the second step  (far zone solution) the solution for large distances downstream of the separatrix characteristics can be obtained  using the approach of \cite{BT}.

\subsection{Method for the solution in the nearest zone} 

The computational method used in this work  was developed for the  solution for the 
problem of the interaction of the relativistic 
pulsar wind with the interstellar medium \citep{koldoba,crab}. 
This method has certain advantages compared to other approaches published in the literature. 
Namely: \\
(i)  The equations which describe the dynamics of the plasma  
are solved only in the shocked region.
There is no need to compute the plasma flow in the pre-shock region as it is done
in previous studies (e.g. \cite{bucciantini,vigelius,swaluw}) 
because the flow in the pre-shock region is \textit{a priory} known; \\
(ii)   All discontinuities are considered as having zero thickness mathematical breaks. 
This allows us to avoid numerical diffusion  of discontinuities. 
This property is especially important for 
calculations of radiation from the shocked material;\\  
(iii) The method has no limitations on the Lorentz factor of the wind  
(e.g. in the approach suggested by \cite{lk} the Lorenz factor 
of the pulsar wind cannot exceed 10). 

We assume that the winds from the pulsar and Be star are isotropic. In this case
flow formed after the interaction  is axially isotropic. In this 
paper we assume that both  winds are cold. Strictly speaking, the stellar wind is hot. 
However, the sound speed in the stellar wind is significantly below 
the bulk motion speed, making this approximation well justified. 
    
The relativistic hydrodynamical (RHD) equations describing the post-shock flow
are  
\begin{equation}
\frac{\partial n \gamma}{\partial t} + \frac{1}{r} \frac{\partial}{\partial r} r n \gamma v_r + 
\frac{\partial}{\partial z} n \gamma v_z =0 
\end{equation}
\begin{equation}
\frac{\partial \gamma^2 w v_r}{\partial t}+ \frac{1}{r} \frac{\partial}{\partial r}r \left( w 
\gamma^2 v_r^2+p \right) +
\frac{\partial}{\partial z}  w \gamma^2 v_r v_z  = \frac{p}{r}
\end{equation}
\begin{equation}
\frac{\partial \gamma^2 w v_z}{\partial t}+ \frac{1}{r} \frac{\partial}{\partial r}rw \gamma^2 v_r 
v_z +
\frac{\partial}{\partial z} \left( w \gamma^2 v_z^2 +p \right) =0
\end{equation}
\begin{equation}
\frac{\partial \left( \gamma^2 w - p \right)}{\partial t}+ \frac{1}{r} \frac{\partial}{\partial r}rw 
\gamma^2 v_r +
\frac{\partial}{\partial z} w \gamma^2 v_z  =0 \ .
\end{equation}
Here  $p$, $w$,$n$ and  $v$ are the pressure, enthalpy, particle density, and
flow velocity, respectively\footnote{Here all thermodynamic quantities have their proper values. $p$ and $w$ are taken per unit volume in the local rest frame.}. At the termination shock front the bulk motion 
energy is transformed  into the energy of the plasma's chaotic motion.  
The efficiency of this transformation depends on the incident angle 
with which plasma crosses the shock wave.  

The plasma flow in the nonrelativistic region  is described  by the following equations:
\begin{equation}
\frac{\partial \rho}{\partial t}+ \frac{1}{r} \frac{\partial r \rho v_r}{\partial r}+
\frac{\partial \rho v_z}{\partial z} =0  \label{eq:nonrel_first}     
\end{equation}
\begin{equation}
\frac{\partial \rho v_r}{\partial t} + \frac{1}{r} \frac{\partial}{\partial r} r \left( \rho v_r^2+p \right) 
+
\frac{\partial}{\partial z} \rho v_r v_z = \frac{p}{r}
\end{equation}
\begin{equation}
\frac{\partial \rho v_z}{\partial t} + \frac{1}{r} \frac{\partial}{\partial r} r \rho v_r v_z +
\frac{\partial}{\partial z} \left( \rho v_z^2+p \right) = 0
\end{equation}
\begin{eqnarray}\nonumber
\frac{\partial}{\partial t} \left( \frac{\rho v^2}{2} + \epsilon \right) + \frac{1}{r} \frac{\partial}{\partial r} r v_r \left( \frac{\rho v^2}{2} + \epsilon +p \right)+ \\
+\frac{\partial}{\partial z} v_z \left( \frac{\rho v^2}{2} + \epsilon +p \right) =0  \label{eq:nonrel_last} 
\end{eqnarray} 
Here $\rho$ and  $\epsilon$ are the  densities of mass and thermal energy, respectively, 
and $p$ ($=2/3 \epsilon$) is the  pressure. 
\begin{figure}
\centerline{\includegraphics[width=0.5\textwidth, bb= 0 0 467 826]{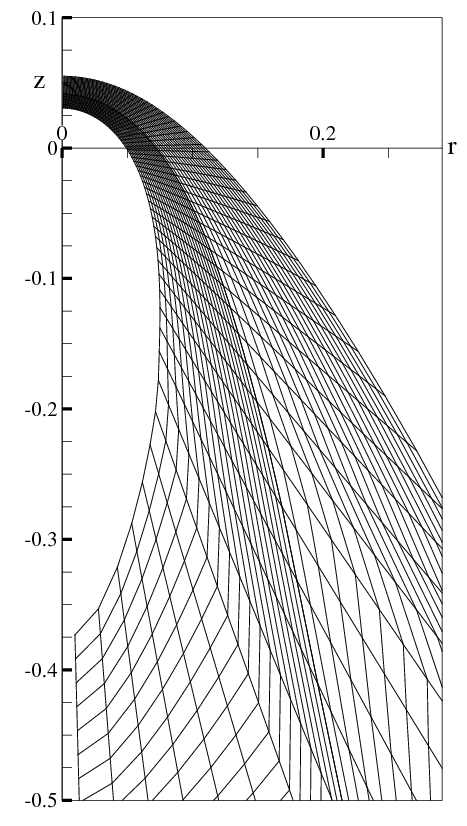}}
\caption{Typical mesh in the computational domain of the nearest zone region. Only every tenth cell is shown.}
\label{mesh}
\end{figure}

Generally, one faces certain difficulties while solving numerically the  
equations which describe interaction of relativistic and nonrelativistic winds. 
The post shock flow is described by the systems of equations which are 
essentially different for the relativistic and nonrelativistic flow domains.  
The locations of  discontinuities (shock waves and the contact discontinuity) 
are unknown \textit{a priori}.  In addition, there are other 
difficulties  related to  large differences in velocities of the 
relativistic and nonrelativistic flows.  The velocity of relativistic plasma 
is equal to $c$ in the pre-shock region and larger than $c/3=10^{10} \ \rm cm/s$ 
in the post shock region. The velocities 
of the nonrelativistic stellar wind in PSR~1259/SS2883 are   $\sim 3\cdot 10^8 \ \rm cm/s$ 
and 
$\sim 10^8 \ \rm cm/s$ in the pre-shock and post-shock regions, respectively.  
The spacial scales in the nonrelativistic and relativistic flows are similar  and have to be  meshed with cells of similar space resolution.  This means that at integration of the full system of equations with explicit numerical schemes,  the time step will be limited by the Courante condition in the region of the relativistic flow.  In the region of the nonrelativistic plasma flow this time-step 
will be less than the Courant limited time-step by two orders of magnitude. 

To overcome all these difficulties,  a special numerical method has been developed for  integration of transient HD and RHD equations.  The integration of the HD equations was performed in the region  limited by the nonrelativistic termination shock wave and by the contact discontinuity.  The RHD equations have been integrated in the region  located between the relativistic termination shock wave and the contact discontinuity.   The plasma flow outside these regions is known. 

An adaptive computational mesh has been used. The mesh boundaries were located on the shock waves and the contact discontinuity. The boundary node location varied  with change of the discontinuity locations.  Usually the nodes were located on the system of \textit{beams} which were fixed. The nodes moved only along these beams.  The beams were  directed along an axis of symmetry of the problem or spread radially from a fixed point (pulsar position for example) as it is shown in Fig.~\ref{mesh}. The nodes on the beams located on the discontinuities were specified. The internal nodes were distributed between them uniformly.      
 
The numerical integration of the HD and RHD equations on the adaptive mesh has been  performed with developed numerical scheme of Godunov type of the second order for spacial variables \citep{godunov}.   The fluxes of  the conservative variables on the cell faces were calculated with a help of the approximate 
solution of the Riemann problem  with the limitation of antidiffusion terms.  In the nonrelativistic domain the Roe linearization \citep{roe}  has been used. 

\begin{figure}
\centerline{\includegraphics[width=0.5\textwidth,bb= 0 0 596 332]{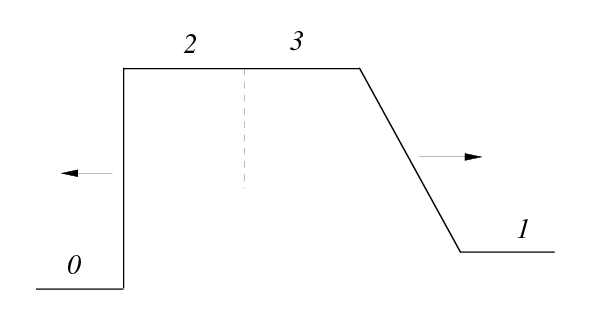}}
\caption{The scheme of the RHD discontinuity decay. The discontinuity can decay into the shock terminating the infalling cold plasma, tangential discontinuity and a shock or rearefication wave propagating in the shocked material.
The case when the  shock propagates in the shocked material is shown. In the linearized case there is no  difference between the shock and  the rearefication wave in the analytical representation.}
\label{decay}
\end{figure}

The location of the  shock fronts and the contact discontinuity have been defined in the process of the solution.  The discontinuity decay has been explored 
for  any discontinuity to find  the location and velocity at the next time step. 
The Riemann problem was solved exactly at the nonrelativistic shock position. 
The Riemann problem at the relativistic shock front was solved in the acoustic approximation.   Schematically the  decay of the discontinuity between the cold ( $p=0$ ) relativistic plasma inflow (state "0") and hot relativistic plasma outflow (state "1") is shown in Fig.~\ref{decay}.  At the calculation of the decay problem it was assumed that:\\
(i)  the termination shock is close to the steady state and the variation of all parameters can be described by linearized ( in relation to the steady state) Gugonio relationships.  \\
(ii)  the variation of the parameters in the post shock region behind any rarefication or shock wave can be described by the linearized equations of RHD.  

\begin{figure}
\centerline{\includegraphics[width=0.5\textwidth, bb=0 0 596 398]{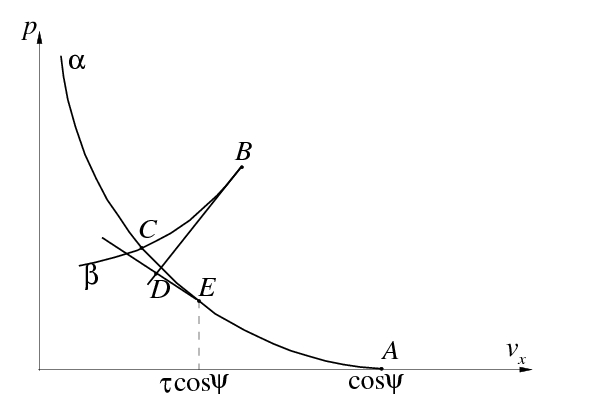}}
\caption{Diagram of RHD discontinuity decay in variables ($p,v$).}
\label{diagram}
\end{figure}

Amplitudes of the waves propagating in the states "0" and the state "1" are taken to 
satisfy the conditions at the contact (tangential) discontinuity: pressure and the 
normal component of the velocity are constant.  The diagram of the RHD discontinuity decay
in variables  ($p,v$) is shown in Fig.~\ref{diagram}. The curve $\alpha$ is the Taube shock adiabata,  going from the point A - initial state of the plasma in the pre-shock region.  Point E corresponds to the case when the termination shock wave is in the steady state.  In the last case  $p=(2Q/3) cos^2 \Psi$ down stream the shock, where $Q$ is energy density in the ultrarelativistic wind, $\Psi$  
is the angle between the flow line and normal to the shock front.  The curve $\beta$  is the shock adiabata going from the point B - the state of the plasma downstream the discontinuity.  The point C, where the curve
$\alpha$ crosses   the curve $\beta$  corresponds to the state at the perfect solution of the Riemann problem.  In our approach the state between the waves corresponds to the point D where the tangents to shock adiabates in the points E and B cross each other.  It follows from the figure that the closer the points E and B placed to each other more accurate are the solution.  

Nonrelativistic HD equations allow a simple scaling of variables: $t \to t/ \alpha$, $\vec v \to \alpha \vec v$, $\rho \to \rho / \alpha^2$, $p \to p$. Moreover, this scaling does not change the conditions  $[p]=0$, $v_n=0$ (which are viable in the steady state regime) at the contact discontinuity. Thus, the difficulties related to large difference in velocities  of the relativistic and nonrelativistic flows may be overcome by rescaling of initial conditions for Eqs.(\ref{eq:nonrel_first})-(\ref{eq:nonrel_last}) in such a way that $v \to c$ in the nonrelativistic wind.  This approach allows us also to stabilize  the flow at the contact discontinuity which in fact is  a tangential discontinuity.

\subsection{Parameterization of the solution}

The steady state equation systems which describe both the nonrelativistic and relativistic outflows can be scaled through the  value of the momentum flux  of the 
nonrelativistic wind, $\rho v^2$, at the distance $D$ between the star and the pulsar. 
Also, it is natural to normalize all geometrical variables to $D$. In all figures below all the 
distances are given in this normalization unless stated otherwise.
With such a normalization   
$\rho v^2=1$ at the dimensionless distance $D=1$.  At the same time the momentum flux of 
the relativistic wind at  $D=1$  becomes 
equal to the parameter $\eta$ given by Eq.(\ref{eq_eta}).  
In this way the interaction of the two flows is described by two parameters: 
$\eta$ and  the initial Lorentz factor of the pulsar wind $\Gamma_0$. 
In the limit $\Gamma_0 \gg1$, the enthalpy per particle $w/n \gg 1$, therefore 
the properties of the flow  only weakly  depend on  the initial 
Lorentz factor $\gamma_0$, thus in the case of pulsar winds with $\Gamma_0 \gg 1$, 
the ratio of the momentum fluxes of the two winds $\eta$ becomes the key parameter which 
describes the system of interacting winds.  
Finally we note that the position (the distance to the pulsar) of the contact discontinuity on the 
symmetry axis may be approximately defined as 
\begin{equation}
r_d={\sqrt{\eta}\over (1+\sqrt{\eta})} \ .
\label{rd}
\end{equation}

\subsection{Method for the solution in the far zone} 

For magnetohydrodynamic nonrelativistic and relativistic flows, 
a method (hereafter BT method) 
has been suggested  by  \cite{BT}.  Below we discuss  a pure
hydrodynamical version of this method which can be applied to the far zone.   

The flow in the far zone is supersonic.
The system of equations describing the steady state flow is hyperbolic.
Therefore, an initial-value Cauchy problem can be formulated for this flow. 

It is convenient to introduce the
flux function $\psi$ as follows

\begin{equation}
n{ \bf u}_{p}={{\bf\nabla}\psi\times \hat {\bf\varphi}\over r},
\label{psi}
\end{equation}
where $\hat {\bf\varphi}$ is the azimuthal unit vector.

The stationary HD equations accommodate the energy conservation law
\begin{equation}
(e+\pi)\gamma=W(\psi) \ ,
\label{eq10}
\end{equation}
where $e$ and $\pi$ are the thermal energy and pressure per particle.
According to this law,  the average energy of particles 
remains constant along the flow line.
This equation should be supplemented by
the relativistic relationship between
the components of the four-velocity $u$
\begin{equation}
\gamma ^{2}=1+u_{p}^{2}.
\end{equation}

For analysing the behavior of the plasma at large distances, it is
convenient to deal with this equation
in an orthogonal curvilinear coordinate system ($\psi, \alpha$) formed by
the flow lines and by the lines perpendicular to them. While 
$\psi$ varies with the motion across the flow lines,
the coordinate $\alpha$ varies with the motion along the flow lines.
The geometrical interval in these coordinates can be expressed as
\begin{equation}
(d{\bf r})^{2}=g^2_{\psi}d\psi^{2}+g^2_{\alpha}d\alpha^{2}+r^{2}d\varphi^{2},
\end{equation}
where $g^2_{\psi}, g^2_{\alpha}$ are the corresponding line elements (components
of the metric tensor).

According to \cite{fields} the equation
$T_{; k}^{\psi k}=0$ (where $T^{ij}$ is the energy-momentum tensor and ";k" denotes covariant
differentiation in these coordinates) can be written in the form
\begin{equation}
{\partial p\over\partial\psi}-
u_{p}^2w{1\over g_{\alpha}} {\partial g_{\alpha} \over \partial\psi}
 = 0.
\label{transfield}
\end{equation}

The unknown variables here
are $z(\alpha, \psi)$ and $r(\alpha, \psi)$.
The metric coefficient $g_{\alpha}$ can be obtained from Eq.(\ref{transfield}),

\begin{equation}
 g_{\alpha} =\exp{\left(\int\limits_0^\psi G(\alpha,\psi)d\psi\right)}\,,
 \label{ga}
 \end{equation}
 where
\begin{equation}
G(\alpha,\psi)={1
\over  u_{p}^2w}{\partial p\over\partial\psi}\,.
\label{G}
\end{equation}
The lower limit of the integration in Eq.(\ref{ga}) is chosen to be 0
such that the coordinate $\alpha$ is uniquely defined.
In this way $\alpha$ coincides with the coordinate $z$ where the
surface of constant $\alpha$ crosses the axis of rotation.

The metric coefficient $g_{\psi}$ can be obtained from  Eq.(\ref{psi})
in terms of the
magnitude of the poloidal velocity 
as follows
\begin{equation}
g_{\psi} ={1\over rn u_{p} }
\,.
\end{equation}

The orthogonality condition
\begin{equation}
 r_{\alpha}r_{\psi}+z_{\alpha}z_{\psi}=0
\,,
\label{orto}
\end{equation}
and the relationships
\begin{equation}
g^2_{\alpha}=r_{\alpha}^{2}+z_{\alpha}^{2} ~~~~ {\rm and} ~~~~ g^2_{\psi}=r_{\psi}^{2}+z_{\psi}^{2}
\label{galpha}
\end{equation}
results in
\begin{equation}
r_{\alpha}=-{z_{\psi} g_{\alpha} \over g_{\psi}} ~~~~ {\rm and} ~~~~ z_{\alpha}={r_{\psi} g_{\alpha}
\over g_{\psi}}
\,,
\label{za}
\end{equation}
with $g_{\alpha}$ given by Eq.(\ref{ga}).
Here $r_{\alpha}=\partial r/\partial\alpha$, $z_{\alpha}=
\partial z/\partial\alpha$, $r_{\psi}=\partial r/\partial\psi$,
$z_{\psi}=\partial z/\partial\psi$.

Eqs.(\ref{za}) should be supplemented by
appropriate boundary conditions and initial values on some initial surface
of constant $\alpha$
located in the nearest zone, but downstream from all the critical surfaces.
The form of the initial surface of constant $\alpha$
was obtained
numerically  via  integration of the following equations

\begin{equation}
{\partial r\over \partial\psi}={nu_{z}\over r(nu_{p})^{2}},~~~~
{\partial z\over \partial\psi}=-{nu_{r}\over r(nu_{p})^{2}}
\,.
\end{equation}
where $nu_z$ and $nu_r$ as well as the integral $W(\psi)$
are given by solution in the nearest zone.

The boundary conditions on the axis of rotation and the equatorial plane are
the same as the conditions in the nearest zone. No conditions at  infinity
are specified.

\begin{figure}
\centerline{\includegraphics[width=0.5\textwidth, bb= 0 0 520 514 ]{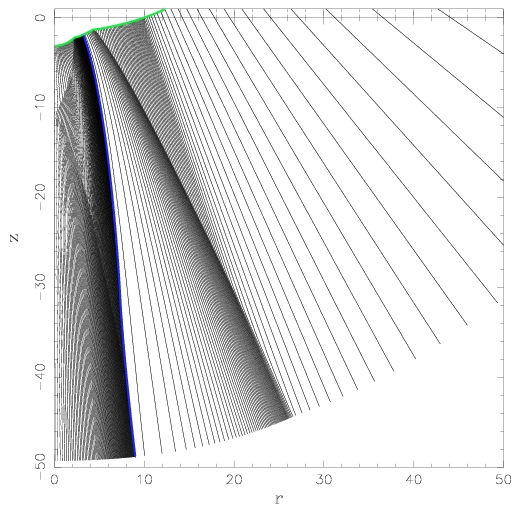}}
\caption{The position of the starting line (green line), flow lines forming the coordinate 
system, contact discontinuity (blue line)  and shock fronts at $\eta=2.5\times 10^{-3}$. 
Only every 10th flow line is shown here. Pulsar is located at the point with coordinates $r=0$, $z=0$. Be star is located at the point with coordinates 
$r=0$, $z=20$. Geometry is normalized by the distance between the pulsar and the terminating relativistic shock wave.}
\label{f11}
\end{figure}

Fig.~\ref{f11} demonstrates a typical structure of the solution and the coordinate system 
obtained  with the BT method. The terminating shocks are well reproduced. 
The nonrelativistic and relativistic flow are perfectly separated. The distribution of the selected flow lines and
correspondingly of the coordinate system lines was chosen to reproduce accurately 
the structure of the shocks.

\section{Results}

The modeling of hydrodynamical collision of two winds 
from a massive star and pulsar has been performed for a wide range of $\eta$ parameter:  
from $\eta= 10^{-4}$, to $\eta=20$, which covers both cases 
when momentum flux is dominated by the wind of pulsar ($\eta > 1$) or by 
optical star  ($\eta < 1$).

\begin{figure}
\centerline{\includegraphics[width=0.5\textwidth, bb=0 0 596 300 ]{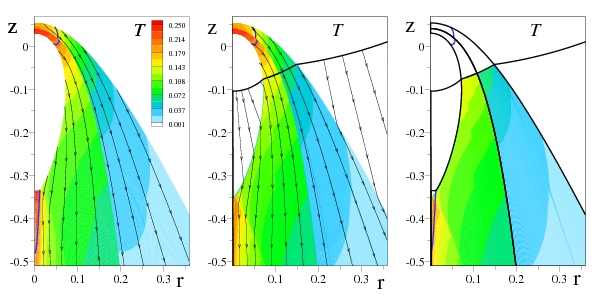}}
\caption{Comparison of the results obtained by different methods in the general computational domain for $\eta=1.1 10^{-3}$. The color represents the distribution of temperature in the post shock region. The left panel shows the solution obtained by the relaxation method (nearest zone solution). The middle panel shows the solution obtained in the far zone and the right panel shows the comparison of the solutions obtained by these methods.The solid lines here show the position of the flow lines and
shock fronts obtained by the relaxation method. Color represents the solution obtained by BT method. In the region of the relativistic flow the temperature is expressed in the units $mc^2u_0$. In the nonrelativistic region the temperature is expressed in the units $mv_0^2$, where $v_0$ is the initial velocity.}
\label{comparison}
\end{figure}

\begin{figure}
\centerline{\includegraphics[width=0.3\textwidth, bb = 0 0 133 168]{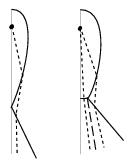}}
\caption{The possible structure of the shock waves at the reflection from the symmetry axis. The thick solid line - the shock fronts. The thin solid line -  flow line. Dashed line - the tangential discontinuity. Two variants are possible. On the left panel the perfect reflection. On the right panel - Mah reflection.}
\label{triple}
\end{figure}

Two computational methods have been applied for solutions  
in the near and far zones. The comparison of two methods in the  far zone 
is shown  in Fig.~\ref{comparison}. Although the two solutions are in good agreement, being practically 
identical, nevertheless  there is a deference between these solutions, namely   
the  shock structures predicted by two methods are  somewhat 
different at the backside point of the pulsar wind termination shock.  
In the  backside point the post-shock supersonic relativistic flow  converges with the axis, thus
a reflecting shock is formed.  The structure of the reflecting shock depends on the angle 
to the symmetry axis. This behavior of the flow is analogous to 
the reflection of the shock wave from a solid wall \citep{gasodynamica}. Two different structures are possible at the reflection: (i) perfect reflection  and 
(ii) Mach reflection. In the last case a triple configuration of the shock front with a tangential discontinuity is formed (see fig \ref{triple}).   In this structure  a region of subsonic flow  is formed. While the  BT method for the  problem solution in the far zone 
is applicable only for pure supersonic flow, 
the subsonic region is not reproduced properly.  However, the subsonic domain occupies a small
region on plots shown in Fig.~\ref{comparison}, otherwise   
the BT solution correctly reproduces all features relevant to 
the interpretation of observations.

\begin{figure}
\centerline{\includegraphics[width=0.5\textwidth, bb= 0 0 362 578 ]{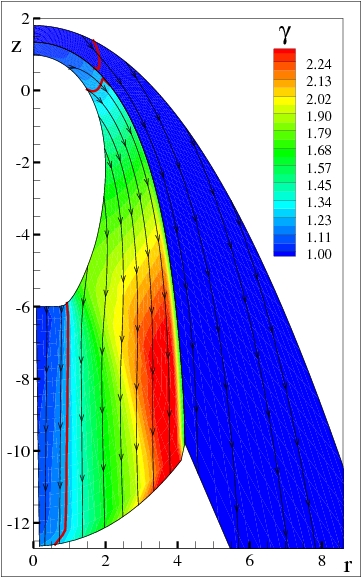}}
\caption{The structure of the post shock flow in the nearest zone solution  at the interaction of the relativistic wind with the plane parallel flow of the nonrelativistic plasma. The color represents the Lorentz factor. Red thick lines show the sound lines. Geometry is normalized on the distance between the pulsar and the relativistic shock front at the symmetry axis.}
\label{parallel}
\end{figure}

\begin{figure}
\centerline{\includegraphics[width=0.5\textwidth, bb= 0 0 580 580 ]{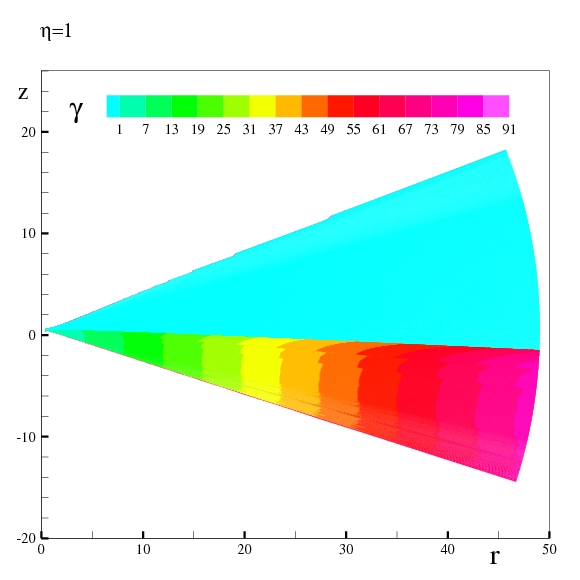}}
\caption{The post shock flow for $\eta=1$. The color represents the bulk Lorentz factor.}
\label{eta1}
\end{figure}

The interactions of the pulsar wind with a  plane parallel nonrelativistic flow 
always result in the formation of a closed termination shock wave 
as shown in Fig.~\ref{parallel}. Similar structure has been 
revealed in calculations by \cite{bucciantini,vigelius,swaluw}. 
However, this assumption cannot be applied to the  radially expanding winds. 
The shock wave associated with the termination of 
the pulsar wind becomes closed  only in the case when
the momentum flux in the nonrelativistic wind significantly  
exceeds the momentum flux in the pulsar wind, namely  for  
$\eta < 1.25\cdot 10^{-2}$. For larger $\eta$  the shock is open  
as shown in Fig.~\ref{eta1}.

\begin{figure}
\centerline{\includegraphics[width=0.5\textwidth, bb = 0 0 553 826]{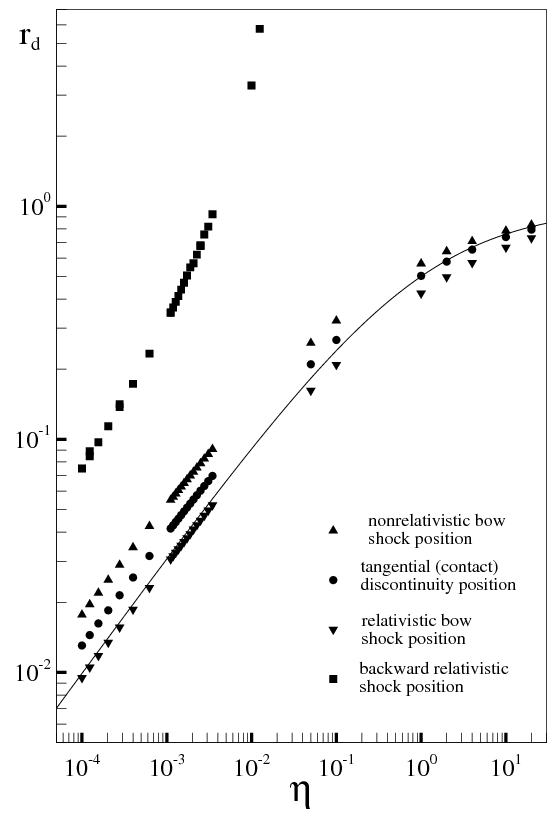}}
\caption{$\eta$ dependence of the location of the shock waves at the symmetry line. The solid line correspoends to the  analitical approximation defined by Eq.(\ref{rd}). }
\label{radius}
\end{figure}

The properties of the post shock flow can be characterized by the location of the  termination
shocks at the symmetry line and by the asymptotical properties of the shocks. 
Dependence of the location of the shocks and the contact discontinuity on $\eta$  at the symmetry line (at the bow shock) is shown  in Fig.  \ref{radius}.  At small $\eta$,
the position of the discontinuity relative to the pulsar   
$r_d \sim \sqrt{\eta}$.   
The distance to the backward point of the relativistic shock front does not follow this law; 
it diverges at $\eta \approx 1.25\cdot 10^{-2}$.

\begin{figure}
\centerline{\includegraphics[width=0.5\textwidth, bb= 0 0 580 324]{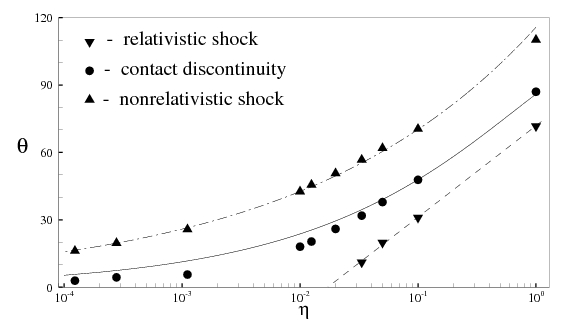}}
\caption{$\eta$ dependence of the  asymptotical opening angle of the shocks and discontinuities in the post shock flow together with interpolations defined by Eqs.(\ref{dd0})--(\ref{dd}). }
\label{angle}
\end{figure}

An important parameter characterizing the post-shock flow is the asymptotic opening angle $\theta$.
Dependence of $\theta$ on $\eta$ is shown  in Fig.~\ref{angle} 
for calculated  positions of 
the nonrelativistic bow shock, the tangential contact discontinuity, and the relativistic bow shock.
The  function $\theta (\eta)$ can be interpolated as  
\begin{equation}
\theta = \exp (4.7516) \eta^{0.217} = 115.7693 \eta^{0.217}
\label{dd0}
\end{equation}
for the nonrelativistic shock wave,  
\begin{equation}
\theta = 41.068 \lg \eta + 71.693
\end{equation}
for the relativistic shock wave, and 
\begin{equation}
\theta = 28.64(2-\eta^{2/5})\eta^{1/3} 
\label{dd}
\end{equation}
for the contact discontinuity.

Interestingly,
Eq. \ref{dd} agrees quite well with the   
analytical interpolation of the asymptotic opening angle
of the contact discontinuity    proposed by \cite{usov}, despite 
the fact that the interpolation of \cite{usov}
is based on the results of simulations of \cite{winds} performed 
for collisions of nonrelativistic winds.

\begin{figure}
\centerline{\includegraphics[width=0.5\textwidth, bb= 0 0 580 580 ]{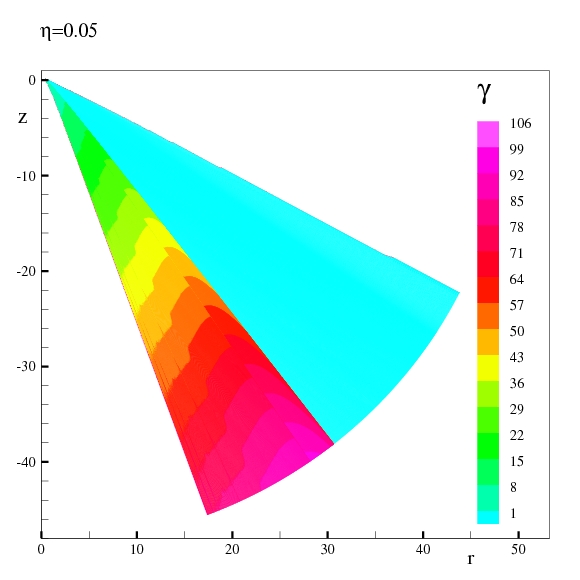}}
\caption{The post shock flow for $\eta=0.05$. The color represents the bulk Lorentz factor.}
\label{eta20}
\end{figure}

It follows from Fig.~\ref{eta1} that the post-shock flow 
is accelerated to rather large Lorentz factors. Even in the nearest zone (limited by the radial distance equal to the distance between the two stars  from
the symmetry axis)  the bulk motion Lorentz factor increases up to $\gamma=2$. In the far zone 
the Lorentz factor becomes very large, e.g. in the zone limited by $r=50$ 
(the size of the computational domain) the bulk Lorentz factor can be as large as 100.  
The results of calculations shown in  Fig.~\ref{eta1} are performed for 
$\eta=1$ which for the system PSR 1250/SS2983 approximately 
corresponds to the case  when the pulsar interacts with the  polar wind of the Be star.  
In Fig.~\ref{eta20} we show also the results for 
$\eta=0.05$ which corresponds to the location of the pulsar in the equatorial wind  
of the Be star.  This case is shown in Fig.~\ref{eta20}. Even in this case, the 
post-shock flow can be accelerated in the downstream region to very large Lorentz factors.

This interesting effect obviously is related to adiabatic losses.
According to  (\ref{eq10}) the full energy per particle,
which  is the product of the bulk motion Lorentz factor $\gamma$ and enthalpy $(e+\pi)$, 
is conserved along the flow line.  Thus the acceleration of the flow
is reduced to the transfer of the thermal energy to the bulk motion.  
Initially  the energy of particles  in the cold pulsar wind 
is domined by the kinetic energy of the motion. 
At the shock the total energy per particle remains unchanged, but 
a part of the energy is transformed to the thermal energy of particles.  

The relationship between bulk motion Lorentz factor and post shock enthalpy 
depends on the incident angle between the shock front and the flow line. 
At the symmetry line, practically all bulk motion energy is transformed 
to the internal energy (enthalpy) of particles. Downstream of the shock 
the pressure reduces because of the flow expansion, and thus 
the bulk motion Lorentz factor increases. Correspondingly the 
temperature falls down, in other words we deal with 
adiabatic losses.  Formally,  the bulk motion Lorentz factor can achieve the initial value 
provided that the post-shock is not terminated by the interstellar medium.

\begin{figure}
\centerline{\includegraphics[width=40mm, bb=0 0 168 579]{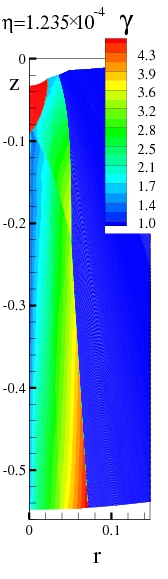}}
\caption{The post shocked flow in the far zone for $\eta=1.2\cdot 10^{-4}$. The color represents the Lorentz factor.}
\label{largeta}
\end{figure}

The bulk motion acceleration to very large  Lorentz factors is possible  not only 
at the collision of the pulsar and stellar winds of comparable power 
as shown in Fig.~\ref{eta1}.  Even in the case when the  momentum flux in the Be 
wind strongly exceeds the momentum flux in the pulsar wind,  
the-post shock flow can be reaccelerated to $\gamma \gg 1$  as it shown in fig. \ref{largeta}

\section{Discussion: Implications and limitations}

In this paper, we have conducted a detailed numerical study of the
interaction of the relativistic and nonrelativistic winds  
assuming isotropic, radially expanding winds and ignoring the role of the magnetic field.
In fact, the cold ultrarelativistic pulsar winds are generally believed 
to be highly anisotropic \citep{bogovalov99,lyubarsky}. The anisotropy of the 
energy flux in the wind can result in a nonaxisymmetric form of the 
termination shock wave at the vicinity of the symmetry axis, as it follows from the 
studies of \cite{vigelius}.  Whether the anisotropy has a noticeable 
impact on  the result and conclusions
of this paper  is a subject of further investigations and will be discussed elsewhere. 
This concerns also the role of the magnetic field. Generally, as it
follows from calculations of the interaction of the pulsar wind with the interstellar medium 
 \citep{crab,lk,bucciantini2}, the role of the magnetic field is rather 
important in the post shock region almost independent of the strength  of the field. 
This is explained by fast amplification of the magnetic field in the post shock region due to 
deceleration of the flow \citep{Khangoulian}.  However, as shown in this paper,
the flow is not decelerated. Just the opposite - it can be accelerated to large bulk motion Lorentz factor.  
Thus although one  should expect an impact  of the magnetic field on the post shocked flow,
this affect most likely will not  be so strong as in the case of the interaction of the 
pulsar wind with interstellar medium.  Detailed MHD calculations 
are needed to to clarify the  role of the magnetic field. 

The effect of reacceleration of the post shock flow to relativistic bulk motion Lorentz factors  
has direct implication to the interpretation of observations of high  energy  $\gamma$-  
and X-rays from  binary pulsar systems like  PSR 1259-63/SS2883.  
This effect strongly modifies the  relationship between the synchrotron X-ray and inverse Compton gamma-ray fluxes 
produced  by the same population of relativistic electrons.   
It is well known that in the  pulsar wind nebulae (plerions) 
which are  formed, by the interaction of pulsar winds with the interstellar medium,
the ratio between  the X-ray  and 
VHE $\gamma$-ray fluxes are defined by the ratio of the energy density of the magnetic field to
the energy density of soft radiation field, provided that  Compton scattering 
takes  place in the Thompson regime \citep{atoyan}.   In binary systems inverse Compton scattering proceeds in the Klein-Nishina regime which
changes the relationship between the  X-ray and VHE gamma-ray fluxes \citep{khangulyan2}.  
Due to our calculations it becomes clear that  a significant deviation from the standard relations  should be  
expected also from hydrodynamics. 

Let us consider these processes in more deatils  assuming that the magnetic field is present in
the wind.
Relativistic particles  moving in the magnetic field  usually produce synchrotron radiation.  
However,  if the wind is cold
 (in this case the velocity of 
particles  coincides with the bulk motion velocity)  these particles do not 
produce  synchrotron radiation. However, they can produce gamma-radiation with
 a specific sharp spectral feature  through the IC scattering  \citep{bogovalov00}.  
An example of such a system is a cold pulsar wind 
which does not produce synchrotron radiation in the pre-shock region. 
In such a system the  ``standart" relation between synchrotron and IC 
radiation components is violated. 
 Remarkably,  the  possibility 
for the formation of relativistic flows in the post-shock region, as revealed in this paper, 
shows that in binary pulsar systems  we should expect a ``non-standard''  
relation between synchrotron and IC 
radiation components in post-shock region as well. 

Moreover, due to the large bulk motion Lorentz factor we should expect strong modulation of 
the observed nonthermal radiation of electrons. Indeed  in the case of 
binary-pulsar systems the direction of the 
post shock flow varies with the motion of the pulsar along the orbit around the  star. 
This implies significant changes of the Doppler factor, $\delta$, 
given the large value of the Lorentz factor. Namely, $\delta \ll 1$ for large viewing angles $\phi$ (e.g. close to $90^\circ$) and
$\delta \gg 1$ for small viewing angles. 
Correspondingly this will have a strong impact on the lightcurve of nonthermal radiation of electrons, $F_\gamma \propto \delta^{n}$ where 
typically $n \geq 3$. This effect, in particular,  can naturally explain the interesting feature of the nonthermal emission of PSR~1259/SS2883, the both synchrotron and inverse Compton components of which 
disappear during the periastron passage of the pulsar (see e.g. \cite{neronov}). This interesting issue will be discussed elsewhere.   

\section*{Acknowledgments}
S.Bogovalov, A.Koldoba and G.Ustyugova are grateful to Max-Planck institute for warm hospitality and financial support during the work on this project.

\end{document}